\documentclass[conference,letterpaper]{IEEEtran}

\usepackage[utf8]{inputenc} 
\usepackage[T1]{fontenc}
\usepackage{url}
\usepackage{ifthen}
\usepackage{cite}
\usepackage[cmex10]{amsmath} % Use the [cmex10] option to ensure complicance
\usepackage{amsfonts,amssymb,mathrsfs,physics,bbm}
\usepackage{stmaryrd} 
\usepackage{ulem}

\DeclareMathOperator{\im}{im}

\DeclareMathOperator\supp{supp}
\usepackage{algorithm}
\usepackage{algorithmicx}
\usepackage{algpseudocode}

\makeatletter
\newcommand\fs@norules{\def\@fs@cfont{\bfseries}\let\@fs@capt\floatc@ruled
  \def\@fs@pre{}%
  \def\@fs@post{}%
  \def\@fs@mid{\kern3pt}%
  \let\@fs@iftopcapt\iftrue}
\makeatother
\floatstyle{norules}
\restylefloat{algorithm}
\usepackage{graphicx}
\usepackage{array}
\ifCLASSOPTIONcompsoc
 \usepackage[caption=false,font=normalsize,labelfont=sf,textfont=sf]{subfig}
\else
 \usepackage[caption=false,font=footnotesize]{subfig}
\fi

\usepackage{xcolor}

\begin{document}
\title{Optimizing hypergraph product codes with random walks, simulated annealing and reinforcement learning}

\author{%
  \IEEEauthorblockN{Bruno Costa Alves Freire}
  \IEEEauthorblockA{Inria Paris, Pasqal\\
                    bruno.costa-alves-freire@inria.fr}
  \and
  \IEEEauthorblockN{Nicolas Delfosse}
  \IEEEauthorblockA{ni.delfosse@gmail.com}
  \and
  \IEEEauthorblockN{Anthony Leverrier}
  \IEEEauthorblockA{Inria Paris\\ 
                    anthony.leverrier@inria.fr}
}

\maketitle

\begin{abstract}
Hypergraph products are quantum low-density parity-check (LDPC) codes constructed from two classical LDPC codes. Although their dimension and distance depend only on the parameters of the underlying classical codes, optimizing their performance against various noise channels remains challenging. This difficulty partly stems from the complexity of decoding in the quantum setting. The standard, \textit{ad hoc} approach typically involves selecting classical LDPC codes with large girth.
In this work, we focus on optimizing performance against the quantum erasure channel. A key advantage of this channel is the existence of an efficient maximum-likelihood decoder, which enables us to employ optimization techniques based on sampling random codes, such as Reinforcement Learning (RL) and Simulated Annealing (SA).
Our results indicate that these techniques improve performance relative to the state-of-the-art. 
\end{abstract}

\section{Introduction}

Quantum Low-Density Parity-Check (qLDPC) codes~\cite{qldpc} are becoming an increasingly attractive approach for the implementation of fault-tolerant quantum computing because their high rate translates into reduced overhead compared to the surface code and its vanishing rate~\cite{gottesman2014fault,constoverheadftqc,tremblay2022constant}. This is achieved by relaxing some hardware constraints and requires long-range connectivity between qubits. Such connectivity might be compatible with architectures based on neutral atoms~\cite{Henriet_2020, xu2023constantoverheadfaulttolerantquantumcomputation}, trapped ions~\cite{monroe2014large, moses2023race, chen2024benchmarking, Hong_2024} and maybe superconducting qubits if one manages to manufacture long-range couplers~\cite{bravyi2024high}.

While specifying a classical LDPC code amounts to selecting a sparse parity-check matrix, specifying a qLDPC code is more complex, as its parity-check matrices must satisfy some additional commutativity constraints.
Hypergraph product (HGP) codes \cite{hgp} offer an elegant and versatile solution: to any full-rank parity-check matrix of a classical LDPC code with parameters $[n,k,d]$, one can associate an HGP code with parameters $\llbracket n^2 + (n-k)^2, k^2, d\rrbracket$. Although better parameters can be achieved asymptotically~\cite{panteleev2022,LZ22,DHL23}, HGP codes have the advantage of producing codes of any length. This raises the question of how to choose the classical code to obtain a quantum code with good performance against some noise model of interest. 

In the classical case, the crucial ingredient allowing for the optimization of LDPC codes was the development of efficient decoding algorithms, based on message-passing, that could be analyzed. Here again, the situation is more complex in the quantum case. Message-passing decoders show poor results due to the presence of unavoidable small cycles in the Tanner graph and to the phenomenon of degeneracy, \textit{i.e.}~the existence of constant-weight errors that act trivially on the code, but can be problematic for the decoder~\cite{poulin2008iterative}. The standard solution is to supplement a message-passing decoder with some classical post-processing step in order to reach a good performance~\cite{panteleev2021degenerate,decodingqldpc}. The large computational cost of these decoders makes them unsuitable to the optimization of HGP codes. For this reason, this optimization has so far been done heuristically, relying on the idea of eliminating small cycles from the Tanner graph of the classical codes, for instance with the Progressive Edge-Growth algorithm~\cite{progedgegrowth}.

We approach this problem by taking inspiration from the classical case and focusing first on the simpler case of the erasure channel. While this channel is mostly a useful toy model in the classical case, it is of practical interest for quantum hardware, either because erasures naturally occur in the system (e.g.~detectable atom or ion loss), or because it is beneficial to engineer qubits that suffer loss more than bit-flips or phase flips~\cite{Wu_2022, Scholl_2023, Kang_2023, Kubica_2023}. The main advantage for us is that the maximum-likelihood decoder for the erasure channel relies on Gaussian elimination, which is efficient and can serve as a tool to compare the performance of various codes. Faster, almost optimal decoders also exist in the case of HGP codes~\cite{Connolly_2024,
gokduman2024erasure,yao2024}. This opens up the possibility to apply machine learning (ML) techniques to optimize qLDPC codes.

Machine learning has already been investigated in the context of quantum error correction, as a way to efficiently explore the space of quantum codes. Reinforcement Learning (RL) approaches require defining a specific reward that an agent will try to optimize, as well as a space of states and actions, which are elementary steps performed in this space. The reward can be the distance of the code or its logical error rate under some noise model, as explored in~\cite{quantumlegorl}. However, the complexity of computing these rewards limits the applicability of the method to short codes only. One can focus on special code families, such as Tensor Network codes, for which the distance can be precomputed. Mauron \textit{et al.}~\cite{tensornetworkrl} exploit a simple and flexible physics-inspired RL framework called Projective Simulation (PS) \cite{briegel2013psai,Melnikov_2017} where the state is given by the generators of the stabilizer of the code and the actions consist of tensor-network operations such as node contractions.
PS can also be applied to explore the family of surface codes~\cite{surfacecodes}. For instance, Nautrup \textit{et al.}~\cite{Nautrup2019optimizingquantum} consider actions corresponding to local deformations of surface codes, with a reward given by the code's performance under the erasure channel, which can be computed in linear time for these codes~\cite{delfosse2016squab}.
Webster \textit{et al.} have proposed optimizing quantum codes using evolutionary algorithms~\cite{webster2024engineering}. They built codes that achieve the best known minimum distance for many parameters $\llbracket n, k \rrbracket$ with $n \leq 20$.
However, the complexity of computing the minimum distance makes this approach difficult to scale up to larger code lengths.

An alternative approach is to define a code via its encoding circuit and actions corresponding to adding a gate to the circuit. For instance, Olle \textit{et al.}~\cite{olle2024simultaneous} exploit the Proximal Policy Optimization to perform such an optimization. However, their approach is limited by the exponential complexity of computing the reward based on the satisfaction of the Knill-Laflamme conditions.

A lesson from these previous attempts is that it is crucial to focus on an efficiently computable reward, and the performance of HGP codes against erasures is a strong candidate. Here, we explore various approaches for the optimization of these codes, including a simple exploration of the state space, simulated annealing and reinforcement learning. We compare the results with the standard approach that relies on Progressive Edge Growth.

In Section \ref{optim_section}, we first review the construction of HGP codes, the erasure channel and the various optimization methods we consider. Section \ref{sec:results} presents and discusses the results and points out further directions of investigation. All the code used to model the optimization problem, implement the algorithms, run the experiments, and generate the plots is available in \cite{brunocaf2024github}.

\section{Optimizing HGP codes} \label{optim_section}

\subsection{Hypergraph product codes}
HGP codes are a special class of quantum codes obtained as the product of two classical codes~\cite{hgp}. Although choosing two distinct constituent codes can be beneficial for specific (biased) error models, we will focus on the case where they coincide. Given an $m\times n$ parity-check matrix $H$, the corresponding HGP is a CSS code~\cite{calderbankshor, steane} with parity-check matrices
\begin{align}
H_X  &= [H \otimes \mathbbm{1}_n | \mathbbm{1}_m \otimes H^T]\\
H_Z & = [\mathbbm{1}_n \otimes H | H^T \otimes \mathbbm{1}_m]
\end{align}
that satisfy the relation $H_X H_Z^T = H H^T + H H^T=0$. 
Assuming that $H$ has full rank equal to $k$, the resulting quantum code encodes $K = k^2$ logical qubits within $N = n^2+m^2 = n^2+(n-k)^2$ physical qubits, and has a distance $d$ equal to the distance of the classical code defined by $H$. The HGP code is LDPC if both matrices $H_X$ and $H_Z$ are sparse, which is automatically the case if $H$ is sparse. 

An error pattern in the quantum case is a pair $(e_X,e_Z) \in \mathbbm{F}_2^N \times \mathbbm{F}_2^N$. Its associated syndrome is $(s_X, s_Z) = (H_X e_X^T, H_Z e_Z^T)$, and the goal of a decoding algorithm is to recover the error, up to a stabilizer element. This means that it succeeds if it returns $(e'_X, e'_Z)$ such that $e_X + e'_X \in R_Z, e_Z +e'_Z \in R_X$, where $R_X$ and $R_Z$ are the spaces generated by the rows of $H_X$ and $H_Z$, respectively. 
An immediate difficulty is that $R_X$ and $R_Z$ contain words of constant weight if the code is LDPC. While they do not affect the encoded information, they are problematic for message-passing decoders. Heuristic decoders have been developed that combine message-passing with some postprocessing \cite{panteleev2021degenerate,decodingqldpc,hillmann2024localized,wolanski2024ambiguity}, but do not lend themselves to a nice analysis. 

We focus on the quantum erasure channel, where each physical qubit is erased independently with probability $p$, and replaced with a maximally mixed qubit. Denoting by $\mathcal{E} \subseteq [N]$ the support of the erasure, it means that the restriction of $(e_X, e_Z)$ to $\mathcal{E} \times \mathcal{E}$ is uniformly distributed over the $4^{|\mathcal{E}|}$ possible values and that $(e_X, e_Z)$ is 0 on the remaining coordinates. The error syndrome and $\mathcal{E}$ are communicated to the decoder. Similarly to the classical case, solving the decoding problem is much simpler for erasures and can be done optimally with Gaussian elimination.

It is \textit{a priori} unclear how to optimize the choice of $H$ in HGP. The standard heuristic approach is to minimize the number of small cycles in its Tanner graph. Note that for a quantum code of $10^3$ qubits, the length $n$ of the classical code is less than 30, and small cycles will inevitably be present if the matrix is chosen at random. In particular, the Progressive Edge-Growth (PEG) algorithm was used to find codes with a large girth~\cite{Connolly_2024}, but it is unknown whether this is the optimal strategy.

\subsection{General methodology: states, actions, cost function}

Inspired by \cite{Nautrup2019optimizingquantum}, we frame our problem as that of searching for a good quantum code inside a collection  referred to as a \textit{state space}, as in the jargon of Reinforcement Learning. This search is realized through the application of \textit{actions}, which \textit{locally} modify the codes, thus obtaining new ones, and effectively traversing the state space through such actions. The quality of the codes is measured by a \textit{cost function}. Finally, an algorithm is needed to optimize the codes under such cost function over the state space.

\subsubsection{State space} \label{sssec:statespace}
The state space $\mathcal{S}$ is the space of Tanner graphs associated to $m \times n$ binary parity-check matrices $H$, with fixed row and column weights.
We restrict ourselves here to the case of \textit{regular} or almost regular codes,  \textit{i.e.}\ matrices $H$ for which all the rows and all the columns have the same weight. Allowing for irregular codes could lead to better performance, however, as well known in the classical case.

\subsubsection{Action space} \label{sec:actionspace}
This is arguably the most important part of the modeling. The state space must be closed under the actions, and ideally it should also be connected. Here, we want the action to preserve the weight conditions on the rows and columns of $H$.
The action we consider is easy to explain in terms of Tanner graphs: one picks two edges and swaps their end-points. Concretely, we take two edges $(u_1, v_1)$, $(u_2, v_2)$ and replace them by the new crossed edges $(u_1, v_2)$, $(u_2, v_1)$. Such action is prone to produce parallel edges, in which case we consider only one instance of each edge for the evaluation of the resulting code.

\subsubsection{Cost function} \label{sssec:costfunction}

We measure the performance of each code through a Monte Carlo estimation of the logical error rate under the quantum erasure channel. Instead of performing the full decoding procedure, we verify for each sampled erasure $\mathcal{E}$ whether it can be corrected based on a criterion from~\cite{delfosse2013upper}, which consists in checking whether there exists $v \in \mathbb{F}_2^N$ such that $\supp{v} \subset \mathcal{E}$ and
\begin{align}\label{eq:correctability_condition}
    v \in \left(\ker{H_X} \setminus \im{H_Z^T}\right) \cup \left(\ker{H_Z} \setminus \im{H_X^T}\right),
\end{align}
implying the existence of a logical error supported within the erasure. In that case, the erasure hides syndrome-compatible corrections that differ by a nontrivial logical operator. For such verification one must solve some linear systems, hence the cost function has $O(N^3)$ complexity. 
In the following, we define as logical error rate the probability that a random erasure (with erasure parameter $p$) admits such a $v$, and therefore cannot be corrected by a maximum-likelihood decoder.

\subsection{Optimization strategies}

\subsubsection{Plain exploration}
The simplest optimization strategy is a naive exploration of the state space, navigated through by the application of actions. The plain exploration starts from a given initial state, and performs a random walk of length $L$. At each step, the algorithm computes the cost function of the current state, as well as the cost function of $N-1$ randomly chosen neighbors, that is, states resulting from the application of an action to the current state. In total, the plain exploration strategy computes the cost function $NL$ times. 

\subsubsection{Simulated annealing}
SA is an iterative, heuristic algorithm that exploits stochasticity to find approximate solutions for optimization problems~\cite{simann}. 
It explores the variable space through a random walk, guided by a temperature scheduling function. At each transition, the algorithm samples a random neighboring point and compares its cost function with that of the current point. If the neighbor's value is better, the transition is accepted. Otherwise it is accepted with a probability that depends on the temperature $T$
\begin{equation} \label{eq:paccept}
P_{\mathrm{accept}} := \exp(- \Delta / T),
\end{equation}
where $\Delta$ is difference between the current value and the neighbor's value. 
We adopted a temperature scheduling of the form
\begin{equation}\label{eq:tempsched}
T = \dfrac{1}{1 + \beta (t / t_{\mathrm{max}})^2},
\end{equation}
where $t$ is the iteration number, $t_{\mathrm{max}}$ is the maximum number of iterations, and $\beta$ is a hyper-parameter.

\subsubsection{Projective simulation}

As an RL framework, Projection Simulation (PS) \cite{briegel2013psai,Melnikov_2017} models the interaction between an agent and an environment responsible for issuing observations of the underlying state, as well as rewards for the actions taken by the agent. 
For each state $s\in \mathcal{S}$, one defines a probability distribution $\alpha_{a,s}$ over the action space $\mathcal{A}_s$ quantifying the conditional probability of taking action $a$ given the current state $s$.
The learning algorithm uses the rewards obtained along the exploration to reinforce actions that lead to the accumulation of such rewards, by updating these probability distributions for all states at each step of the exploration. 

The concrete implementation of the PS agent consists of two matrices, $h, g \in \mathbb{R}^{\abs{\mathcal{S}} \times \abs{\mathcal{A}}}$. Each row in $h$ holds the local policy of a state, which is converted into a probability distribution via a softmax transformation $\alpha_{a,s} \propto e^{\beta h_{a,s}}$ where $\beta$ is a hyper-parameter of the model. In the initialization, the $h$ matrix is set to zero,  \textit{i.e.}\, all the distributions $\alpha$ are uniform. 
The glow matrix $g$, initially set to 0, is used to update the $h$ matrix upon reception of a reward $\lambda$ through $ h \gets h + \lambda  g$.

The PS agent also possesses two \textit{forgetting} mechanisms, controlled by the forgetting parameter $\gamma \in [0, 1]$ and the glow damping parameter $\eta \in [0, 1]$. They respectively act on $h$ and $g$ by shrinking their weights: $h \gets (1-\gamma) h$, $g \gets (1-\eta) g$.
The forgetting update takes place right before the incorporation of a reward, so the full update of the $h$ matrix at time $t$ takes the form
\begin{equation}
    h^{(t+1)}_{a,s} = (1-\gamma) h^{(t)}_{a,s} + \lambda^{(t)}  g^{(t)}_{a,s}. 
\end{equation}
In addition to the damping mechanism, one entry of the glow matrix is set to $1$ right after an action is taken: 
\begin{equation}
    g^{(t+1)}_{i j} = \max\left((1-\eta) g^{(t)}_{i j}, \mathbf{1}[s^{(t)} = s_i] \mathbf{1}[a^{(t)} = a_j] \right). 
\end{equation}
A training episode takes hyper-parameters $\gamma$, $\eta$ and $\beta$, and initial state $s^{(0)}$.
The full training may comprise several episodes, resetting the state to the same initial state $s^{(0)}$ but keeping the $h$ and $g$ weights across episodes.

\subsubsection{Implementation details}
We relied on the M4RI library \cite{M4RI} to perform the linear algebraic operations over $\mathbbm{F}_2$ involved in verifying condition \eqref{eq:correctability_condition} for the cost functions. Python implementations of the PS framework and some generalizations are available at \cite{PSmelnikov}. We made use of the NetworkX \cite{networkx} library, as well as other more commonplace Python libraries, such as NumPy \cite{numpy} and SciPy \cite{scipy}.

\section{Results and discussion}
\label{sec:results}
We focused on 3 choices of code parameters, allowing us to compare our optimized codes with those from Connolly \textit{et al.} obtained with the PEG algorithm~\cite{Connolly_2024}. The classical codes are $(3,4)$-LDPC codes, meaning that the columns and rows of the matrix $H$ have weight 3 and 4, respectively. We consider classical codes of parameters $[20,5], [32,8]$ and $[36,9]$ leading to HGP codes with parameters $\llbracket 625,25\rrbracket, \llbracket 1600,64\rrbracket$ and $\llbracket 2025,81\rrbracket$, with a rate of $4\%$. For each of the optimization procedures, the initial state $s^{(0)}$ is the code reported in Connolly \textit{et al.}~\cite{Connolly_2024}.

In all our experiments, we ran $10^4$ Monte Carlo trials to estimate the probability that a code fails at correcting erasures for a given erasure probability $p$. The choice of $p=9/32$ for the codes of length 625 and 1600, and $p=12/32$ for the code of length 2025 was made to yield a ratio between standard error and mean of at most $5\%$.

Table~\ref{tab:explorationparams} reports the best values of the hyper-parameters that were used during the optimization. For SA, we considered the cost function in log-scale, swept the value of $\beta$ through $1, 4, 7, 10$ and indicate the best value. We found that the best codes discovered by SA are often reached in the middle of the temperature schedule, rather than at the end.

\begin{table}[ht]
    \caption{Parameters for the optimization}
    \label{tab:explorationparams}
    \centering
    \begin{tabular}{c|c|ccc}
           &  & $[[625, 25]]$ & $[[1600,64]]$  & $[[2025, 81]]$ \\
         \hline
     &   eras. rate & $9/32$ & $9/32$ & $12/32$\\
	\hline
         plain & $N$ & $24$  & $12$  & $8$ \\
         explor. & $L$ & $120$ & $40$  & $30$ \\
         \hline
         SA & $t_{\mathrm{max}}$ & $2400$ & $450$  & $180$ \\
         & $\beta$  & 4 & 10  & 1 \\
         \hline
         RL &  $\#$ ep. & $20$ & $8$  & $5$ \\
         & max steps/ep. & $120$ & $50$  & $35$ \\
         \hline
         \multicolumn{5}{c}{Hard threshold} \\
         \hline
         RL & $\theta_{\mathrm{hard}}$ & $10^{-2}$ &  $2\cdot 10^{-3}$  & $3\cdot10^{-2}$ \\
         &  $\beta$ & $6.79$ & $9.12$  & $7.97$ \\
         & $\gamma \, (\times 10^4)$ & $4.56$ & $1.72$  & $28.6$ \\
         & $\eta \, (\times 10^3)$ & $1.90$ & $1.58$  & $2.60$ \\
         \hline
         \multicolumn{5}{c}{Easy threshold} \\
         \hline
         RL &$\theta_{\mathrm{easy}}$ & $2\cdot10^{-2}$ & $4\cdot10^{-3}$ & $4\cdot10^{-2}$ \\ 
         & $\beta$ & $6.79$ & $8.43$ &  $8.66$ \\
         & $\gamma$ & $8.98\cdot10^{-4}$ & $0$  & $0$ \\
         & $\eta$ & $0$ & $0$ &  $0$ \\
                  \hline
    \end{tabular}
\end{table}

For Projective Simulation, we set a number of episodes in the training and a maximum number of steps per episode. Similarly to Nautrup \textit{et al.}~\cite{Nautrup2019optimizingquantum}, we adopted a termination criterion for the episodes based on the first emission of a nonzero reward, and a sparse reward function based on a threshold $\theta$, \textit{i.e.}\ the reward is $\lambda = 1$ if the failure probability of the decoder is below $\theta$ and $\lambda = 0$, otherwise. 
For each code family, we picked two values for the threshold, $\theta_{\mathrm{easy}}$ and $\theta_{\mathrm{hard}}$, and three values of $\beta$ (softmax function), $\gamma$ (forgetting mechanism) and $\eta$ (glow damping). For the ``easy'' threshold, the simulation reaches a point where the episodes are terminated in just one step, indicating that the agent has learned how to achieve the desired threshold with little exploration, and thus limiting the best performance. On the other hand, for the ``hard'' threshold, the agent may not receive enough rewards and therefore starve on feedback to learn from.

The best combination of parameters for each code and method is reported in Table~\ref{tab:explorationparams}. It corresponds to the best value of the cost function found among all the codes explored during the optimization procedure. It should be noted that this is a stochastic quantity, which carries a large uncertainty when the logical error rate of the codes under consideration is too low for the number of Monte Carlo trials. Thus, the criterion to choose the optimal parameters from Table~\ref{tab:explorationparams} is a proxy of the codes' performance, but lacks robustness. 
In order to robustly assess the quality of the codes found by each method, we need more precise simulations, and over a broader range of noise rates. Indeed, the codes are selected based on their performance for one fixed noise rate $p$, which is not practically relevant, and only serves the purpose of being large enough to make the cost function easy to estimate with a reasonable number of Monte Carlo samples, and small enough so that the cost function provides the agent with some insights on the code performance in the regime of low logical error rate, which is the regime of practical interest. 

\begin{figure*}[htbp]
    \centering
    \includegraphics[width=0.94\linewidth]{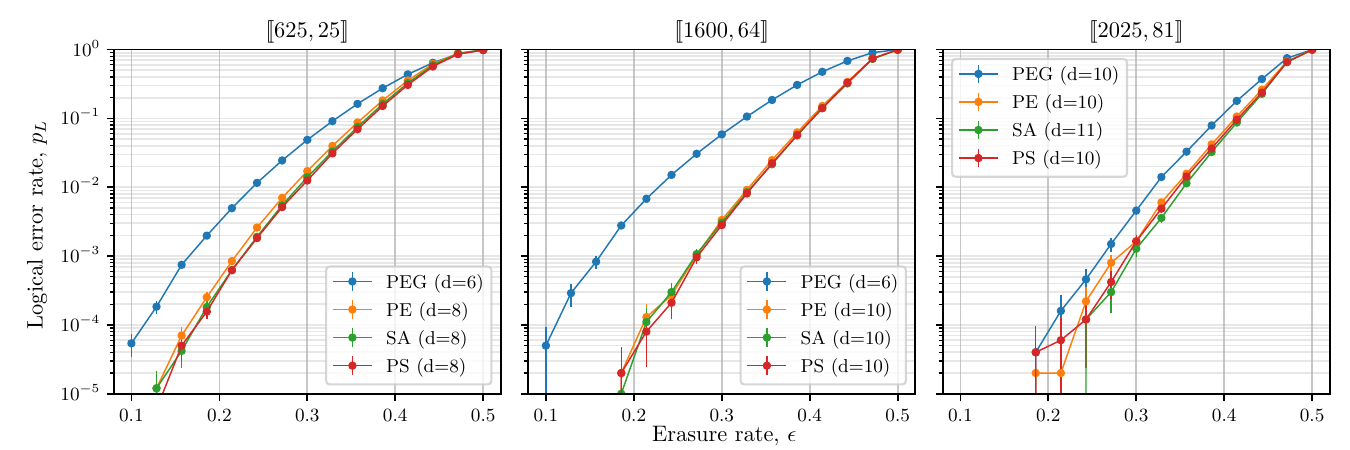}
    \caption{Comparison of the performance under the erasure channel of the original code from \cite{Connolly_2024} (PEG), with the best codes obtained through plain exploration (PE), simulated annealing (SA) and projective simulation (PS), estimated with $5\cdot10^{5}$, $10^{5}$ and $5\cdot10^{4}$ Monte Carlo samples for the respective code lengths. All PEG codes have girth $6$, whereas all optimized codes have girth $4$, which suggests that maximizing girth is not a good heuristics for improving performance.}
    \label{fig:codes_performance}
\end{figure*}

Fig.~\ref{fig:codes_performance} shows the performance over the erasure channel with maximum likelihood decoding of the codes obtained from the different optimization strategies.
While each code family was optimized for a single value of the erasure channel (indicated in Table \ref{tab:explorationparams}), the simulation results show that the optimized codes with length 625 and 1600 outperform the initial ones over a whole range of parameters, and in particular in the regime of low logical error rate. 
This is explained by their resulting code distances, annotated in the plots.

\begin{figure*}
    \centering
    \includegraphics[width=0.94\linewidth]{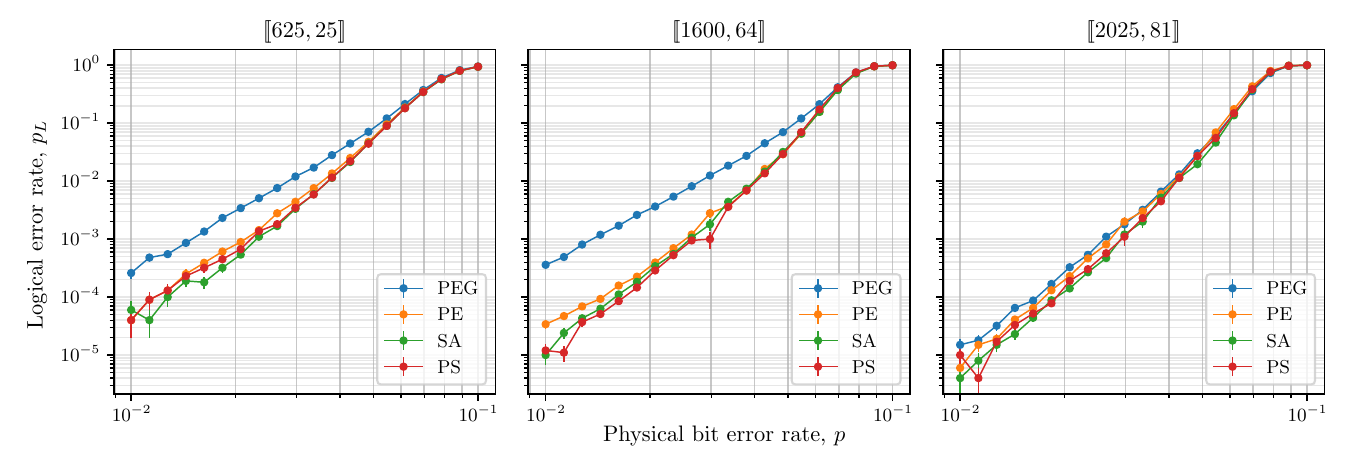}
    \caption{Logical error rate of the codes optimized with PEG \cite{Connolly_2024}, PE, SA and PS, for the bit-flip error channel with physical error probability $p$. The codes are decoded with BP+OSD \cite{panteleev2021degenerate,decodingqldpc}, with the implementation from \cite{Roffe_LDPC_Python_tools_2022}.}
    \label{fig:bposd}
\end{figure*}

Fig.~\ref{fig:bposd} shows that, although trained over the erasure channel, these codes also perform better than the original codes of~\cite{Connolly_2024} over the bit-flip error channel. This type of \textit{transfer learning} is particularly useful because the particular noise that will be present in the lab will never be fully specified, and it is therefore desirable that the optimization procedure provides codes that are robust against variations of the noise model.  
The performance of the length-2025 code is not significantly improved. More extensive numerical simulations may be necessary to optimize such large-length codes.

It is noteworthy that the plain exploration already yields improvements on the initial PEG codes, with comparable performance to SA and PS, which serves to show that the exploration aspect is the key ingredient in all the optimization procedures studied here. 

\section{Conclusion}
In this paper, we demonstrated that several optimization techniques including Simulated Annealing and Reinforcement Learning can be employed to find hypergraph product codes that outperform codes obtained through Progressive Edge-Growth.
Surprisingly, the three optimization strategies we considered lead to very similar performance, suggesting that they may have found close-to-optimal solutions for these code lengths. Furthermore, the codes optimized for the erasure channel also perform well against bit-flips, indicating that performance against erasures might be a good proxy for performance against more realistic noise models.

While the plain exploration and SA are simpler to implement and tune, the main advantage of RL is that it produces a policy capable of exploring the code space, rather than just providing a single optimized solution or path in that space.

This policy can be reused for related optimization tasks, a form of transfer learning illustrated in~\cite{Nautrup2019optimizingquantum}, where agents trained to optimize codes under a specific noise model successfully transferred their learning to other settings.
One can also imagine training an RL agent to locally optimize the Tanner graph of a small code and later applying the resulting policy to much larger codes.

It would be interesting to replace PS with other RL algorithms, particularly actor-critic algorithms~\cite{6392457}. A pair of graph neural networks -- one for proposing actions and another one for estimating rewards -- may reduce state evaluation costs. This could also address a limitation of PS, namely that it learns only from visited states and gains no information about unvisited ones.

Other applications of RL in quantum error correction are possible, such as the design of decoders for surface codes~\cite{sweke2020reinforcement, sivak2024optimization} or quantum LDPC codes~\cite{maan2024machine}.

\section*{Acknowledgment}

We acknowledge funding from the Plan France 2030 through the project ANR-22-PETQ-0006.
Nicolas Delfosse is now with IonQ Inc. His contributions to this work were done before joining IonQ.

\bibliographystyle{IEEEtran}
\bibliography{main}

\end{document}